\documentclass[12pt]{article}

\newcommand{\beq}{\begin{equation}}
\newcommand{\eeq}{\end{equation}}
\newcommand{\beqa}{\begin{eqnarray}}
\newcommand{\eeqa}{\end{eqnarray}}
\newcommand{\beqar}{\begin{eqnarray*}}
\newcommand{\eeqar}{\end{eqnarray*}}

\newcommand{\Ga}{\Gamma}

\newcommand{\inn}{\!\cdot\!}

\newcommand{\la}{\lambda}

\newcommand{\z}{\zeta}

\newcommand{\eg}{{\it e.g.,}\ }
\newcommand{\ie}{{\it i.e.,}\ }
\newcommand{\labell}[1]{\label{#1}} 
\newcommand{\reef}[1]{(\ref{#1})}
\newcommand\prt{\partial}

\newcommand\cR{{\cal R}}

\newcommand\Tr{{\rm Tr}}

\begin{document}

\begin{titlepage}

\begin{center}



{\LARGE \bf Disk level S-matrix elements \\ 
\vskip 0.2 cm
at eikonal Regge limit
 }\\
\vskip 1.25 cm
 Mohammad R. Garousi\footnote{garousi@mail.ipm.ir}  \\
 \vskip 1cm
\vskip 1 cm
{{\it Department of Physics, Ferdowsi University of Mashhad\\}{\it P.O. Box 1436, Mashhad, Iran}\\}
\vskip .1 cm
\vskip .1 cm

\end{center}

\vskip 0.5 cm

\begin{abstract}
\baselineskip=18pt
 We examine the calculation of  the color-ordered  disk level  S-matrix element of  massless  scalar   vertex operators for the  special case that some of  the   Mandelstam variables for which there are no open string channel in the amplitude, are set to zero.   By explicit calculation we  show that the  string form factors in the 2n-point functions reduce to one  at the eikonal Regge limit. 
\end{abstract}
\vskip 3 cm
\begin{center}
\end{center}
\end{titlepage}

\section{Introduction}
The study of scattering amplitudes has been an active subject over the past few years. In particular, it has been shown by Britto, Cachazo, Feng and Witten (BCFW) \cite{Britto:2004ap,Britto:2005fq} that any $n$-point on-shell scattering amplitude can be written as sums of products of lower-point on-shell amplitudes. The BCFW recursion relation has been found by the observation that a tree-level $n$-point scattering amplitude is a rational function of the external momenta. Hence, by analytic continuation of the external momenta into the complex plane, the amplitude becomes meromorphic function which is then uniquely determined by its singularities, \ie its poles and residues.

A meromorphic functions is uniquely determined by its singularities provided that there is no pole at infinity. Hence,  the BCFW recursion relations are valid when the complex scattering amplitude has no pole at large complex parameter $z$. Naive power counting of individual Feynman diagrams seems  to lead to dangerously high power of $z$, however, cancellation among them may result in a much lower power of $z$. A criterion has been found in \cite{ArkaniHamed:2008yf,Cheung:2008dn} which allows one to conclude which theories allow the BCFW recursion relations.  

The applicability  of the BCFW recursion relations to string theory has been studied  in \cite{Boels:2008fc,Cheung:2010vn}. It has been shown explicitly in \cite{Boels:2008fc} that the Veneziano amplitude  has no pole at infinity. Using the pomeron vertex operators \cite{Brower:2006ea}, it has been shown in \cite{Cheung:2010vn} that all tree-level complex string amplitudes lake a pole at large  $z$. This happens in a  particular unphysical kinematic region. In this region then the string BCFW recursion relations hold. The same relations are  then valid   by analytic continuation to the physical region. The  string BCFW recursion relation has been then found explicitly in \cite{Cheung:2010vn} for the scattering amplitude of external open string tachyons. The string BCFW recursion relation for the scattering amplitudes involving both open and closed string has been studied in \cite{Fotopoulos:2010cm}.

It has been observed in \cite{Cheung:2010vn} that the leading and subleading asymptotic behavior of string amplitudes is the same as the asymptotic behavior of their low energy field theory. This leads the authors to conjecture that in a particular limit ( the eikonal Regge (ER) limit) in which some of the kinematic variables are much larger than the string scale and the rest much smaller, the string S-matrix elements are reproduced by the corresponding S-matrix elements in the low energy field theory. This conjecture has been checked explicitly in \cite{Cheung:2010vn} by demonstrating that MHV amplitudes in type I string theory and ${\cal N}=4$ super Yang-Mills theory are in fact equal in this limit at four and five point functions. In this paper we would like to check this conjecture at the six- and higher-point function.

The six-point function has been studied in \cite{Oprisa:2005wu,Stieberger:2006bh} where it has been shown that the amplitude   involves  the complicated triple Hypergeometric function. The amplitude in general is a function of 9 independent  Mandelstam variables. However, to study the amplitude at the ER limit one needs to keep nonzero the Mandelstam variables that are either large at ER limit or they appear as massless pole in the scattering amplitude. We will see that these criteria allows one to set to zero three Mandelstam variables and hence the amplitude become much more easier to study. We will see that the string form factors  for many terms in the scattering amplitude can be written in terms of only the Gamma functions which may then  be easily studied at the ER limit. Some terms in the amplitude can be written in terms of the multiple of the Gamma functions and the triple Hypergeometric functions which are hard to study at the ER limit. In those cases, one may constraint another set of three Mandelstam variables to zero to write the result in terms of only the Gamma functions. In the eight-point functions, we will see that 10 Mandelstam variables can be set to zero which simplifies greatly the calculation of the amplitude. In the general case of 2n-point functions, one can set $(2n-3)(2n-4)/2$ Mandelstam variables to zero. Many terms of the amplitude can be calculated explicitly, and the final results are  in terms of multiple of the  Gamma functions.  

In the next section we review the four-point function and its behavior at the ER limit. In section 3, we calculate the six point function at the particular case that three of the Mandelstam variables are zero. We consider only  terms in  which  the polarization of the gluons contract with each other. This part of the amplitude is the same as the scattering amplitude of six massless scalar fields. There are 15  contractions between the scalar polarizations. We will see that the string form factor in 7 of them are just involve the Gamma function, 4 of them involve the Gamma and the Hypergeometric function ${}_3F_2$, 3 of them involve the Gamma and the Hypergeometric function ${}_4F_3$ and the last one involve the more complicated triple Hypergeometric function. In section 3.1, we show that the form factors for the first 11 terms  are easily reduce to one in the ER limit. For other terms that involve ${}_4F_3$ and the triple Hypergeometric function, one may choose another constraint to write them in terms of only the Gamma functions which are then reduce to one in the ER limit. In section 4, we perform the calculation of 8-point function for the special case of adjacent contractions of the scalar polarizations and show that the form factor reduces to one in the ER limit. In section 5, we extend the result of section 4 to the general case of 2n-point functions.
     
\section{Four-point functions}

It has been shown in \cite{Cheung:2010vn} that the S-matrix element of four and five gauge bosons which can be written in terms  of the  MHV amplitudes are  reduced to the corresponding MHV amplitudes in the Yang-Mills theory at the ER limit. Here we review the argument for the four gauge bosons.

The color-ordered scattering amplitude of four gauge bosons on D-brane is given by \cite{Schwarz:1982jn, Garousi:1996ad}
\beqa
{\cal A}&\sim& K(\z_1,\z_2,\z_3,\z_4)\Tr(\la_1\la_2\la_3\la_4)\frac{\Gamma(2\alpha'k_1\inn k_4)\Gamma(2\alpha'k_1\inn k_2)}{\Gamma(1+2\alpha'k_1\inn k_4+2\alpha'k_1\inn k_2)}
\eeqa
where the kinematic factor is 
\beqa
K&=&-4\alpha'^2k_1\inn k_2(\z_1\inn k_4\z_3\inn k_2\z_2\inn\z_4+\z_2\inn k_3\z_4\inn k_1\z_1\inn\z_3+\z_1\inn k_3\z_4\inn k_2\z_2\inn\z_3+\z_2\inn k_4\z_3\inn k_1\z_1\inn\z_4)\nonumber\\
&&-4\alpha'^2k_2\inn k_3 k_2\inn k_4 \z_1\inn \z_2\z_3\inn\z_4+\{1,2,3,4\rightarrow 1,3,2,4\}+\{1,2,3,4\rightarrow 1,4,3,2\}\labell{kin}
\eeqa
In four dimensions, using the spinor-helicity formalism  \cite{Dixon:1996wi}, the amplitude simplifies to just one term. We are not going to use this formalism in this paper. Using the property of the Gamma function, $x\Gamma(x)=\Gamma(x+1)$, one can write the amplitude as
\beqa
{\cal A}&\sim& {\cal A}_{YM} F
\eeqa
where the field theory amplitude is
\beqa
{\cal A}_{YM}&=&\frac{K(\z_1,\z_2,\z_3,\z_4)}{4\alpha'^2k_1\inn k_4 k_1\inn k_2}\Tr(\la_1\la_2\la_3\la_4)
\eeqa
and the string form factor is 
\beqa
F&=&\frac{\Gamma(1+2\alpha'k_1\inn k_4)\Gamma(1+2\alpha'k_1\inn k_2)}{\Gamma(1+2\alpha'k_1\inn k_4+2\alpha'k_1\inn k_2)}
\eeqa
At low energy, $\alpha' k_i\inn k_j\rightarrow 0$, the string form factor reduces to one. 

The adjacent BCFW shifts is
\beqa
k_i&\rightarrow &\hat{k}_i=k_i+qz\nonumber\\
k_{i+1}&\rightarrow &\hat{k}_{i+1}=k_{i+1}-qz
\eeqa
In order to keep the on-shell conditions,  $q$ must satisfy the relations  $q\inn q=k_{i}\inn q=k_{i+1}\inn q=0$. At the ER limit, one takes $\sqrt{\alpha'}k_j\sim {\cal O}(\epsilon)$ for all $j$ where $\epsilon$ is a small number, $\sqrt{\alpha'} q\sim {\cal O}(\epsilon^{-1})$ so that $q\inn k_j\sim 1$, and then takes $z\rightarrow \infty$. In this limit then  $\alpha'\hat{k}_i\inn k_j$ is large and $\alpha'\hat{k}_i\inn\hat{k}_j,\, \alpha'k_i\inn k_j$ are small. This reduces the form factor to one. However, for non-adjacent BCFW shift, \eg $\hat{k}_1,\, \hat{k}_3$, the form factor does not reduce to one.

There are two different terms in the kinematic factor \reef{kin}. Terms in which the polarizations are contract only with each other and terms in which some of the polarizations are contract with momentum. The Yang-Mills theory produce all of them. In the six-point function there are too many terms of the second type and there are 15 terms of the first type. The string form factors for both type of terms are similar, so for ease of calculation we consider only terms of the first type. 
On the other hand the S-matrix element of these terms are the same as the S-matrix element of six transverse scalar vertex operators. So in this paper we only calculate the S-matrix elements of the scalar fields.

\section{Six-point functions}

The S-matrix element of six gauge bosons for arbitrary Mandelstam variables has been calculated in \cite{Oprisa:2005wu}. The result is in terms of triple Hypergeometric functions. However, we will show that the complicated triple Hypergeometric functions do not appear in the amplitude if one sets to zero three of the Mandelstam variables .  In this case the amplitude can be written in terms of only Gamma function. So let us  calculate the amplitude for six scalar fields.

The color-ordered disk level S-matrix element of six scalar vertex operators is given by the following correlation function:
\beqa
A&\sim &\int dx_1\cdots dx_6<\prod_{i=1}^6V_i(2k_i,x_i)>\Tr(\la_1\la_2\la_3\la_4\la_5\la_6)\labell{amp}
\eeqa
The position of the vertices are $-\infty<x_1<x_2<x_3<x_4<x_5<x_6<\infty$. Since the background charge of the world-sheet with topology of a disk is $Q_{\phi}=2$ one has to choose two of the vertex operators to be in -1 picture and the rest to be in 0 picture. These vertex operators are
\beqa
V^{-1}_i(2k_i,x_i)&=&:\z_i\inn e^{-\phi}\psi e^{2k_i\inn X}:\nonumber\\
V^{0}_i(2k_i,x_i)&=&:\z_i\inn(\prt X+2ik_i\inn\psi\psi)e^{2k_i\inn X}:
\eeqa
where $\z_i$ is the polarization of the scalar fields in the transverse space. Using the standard world-sheet propagators
\beqa
<X^{\mu}(x)X^{\nu}(y)>&=&-\frac{\alpha'}{2}\eta^{\mu\nu}\log(x-y)\nonumber\\
<\psi^{\mu}(x)\psi^{\nu}(y)>&=&-\frac{\alpha'}{2}\frac{\eta^{\mu\nu}}{x-y}\nonumber\\
<\phi(x)\phi(y)>&=&-\log(x-y)
\eeqa
one can calculate the correlators in \reef{amp}. The result should be invariant under $SL(2,R)$, the conformal symmetry of the disk. Removing this  symmetry  by fixing the position of three vertex operators, one would find a triple integral which in general can be written in terms of the triple Hypergeometric functions. We will see that the triple integral simplifies greatly  if one sets three of the Mandelstam variables to zero, \eg 
\beqa
k_2\inn k_4=k_2\inn k_5=k_3\inn k_5=0\labell{zero}
\eeqa
or any other set under cyclic permutation of $(1,2,3,4,5,6)$. Since the amplitude has no  pole in $(k_2+k_4)^2$, $(k_2+k_5)^2$ and $( k_3+k_5)^2$ channels, we are allowed to restrict the Mandelstam variables to the above values. Moreover, for the adjacent BCFW shift $\hat{k}_6, \hat{k}_1$,  the ER limit takes  $\alpha'k_2\inn k_4,  \alpha'k_2\inn k_5,\alpha'k_3\inn k_5\rightarrow 0$ which is consistent with the above constraint.

Under the restriction \reef{zero}, one can set to one the expression $x_{24}^{2\alpha'k_2\inn k_4}x_{25}^{2\alpha'k_2\inn k_5}x_{35}^{2\alpha'k_3\inn k_5}$ which results from the correlation of  $\prod_{i=1}^6e^{2ik_i\inn X}$ in the amplitude \reef{amp}. This simplifies the triple integral that one finds at the end. Moreover, the calculation become more  easier, \ie the final result is in terms of only the Gamma functions,  when  the correlator of the other parts of the amplitude does not produce terms like $x_{24}^{-n_{24}}x_{25}^{-n_{25}}x_{35}^{-n_{35}}$ where $n_{ij}$ are some integer number. This happens when one  restricts  the contraction of the polarization of the scalars to those which does not include 
\beqa
\z_2\inn\z_4,\,\z_2\inn\z_5,\,\z_3\inn\z_5\labell{zero1}
\eeqa
The final result for amplitude involving the above contractions will be  in terms of more complicated functions. 

We choose the following pictures for the vertex operators in \reef{amp}:
\beqa
<V^{0}(2k_1,x_1)V^{-1}(2k_2,x_2)V^{-1}(2k_3,x_3)V^{0}(2k_4,x_4)V^{0}(2k_5,x_5)V^{0}(2k_6,x_6)>\labell{amp1}
\eeqa
Since the scalar polarizations are in the normal space and momenta are in the world volume space, one can perform the correlator of the exponential factors and write the amplitude \reef{amp} in the following form:
\beqa
A&\sim&\int dx_1\cdots dx_6\,\z_{1i}\z_{2j}\z_{3k}\z_{4l}\z_{5m}\z_{6n}\Tr(\la_1\la_2\la_3\la_4\la_5\la_6)x_{23}^{-1}\prod_{i<j}^6x_{ij}^{-s_{ij}}\labell{amp2}\\
&&<(\prt X^i+2ik_1\inn\psi\psi^i)\psi^j\psi^k(\prt X^l+2ik_4\inn\psi\psi^l)(\prt X^m+2ik_5\inn\psi\psi^m)(\prt X^n+2ik_6\inn\psi\psi^n)>\nonumber
\eeqa
where $x_{ij}=x_i-x_j$ and $s_{ij}=-\alpha'(k_i+k_j)^2=-2\alpha'k_i\inn k_j$ are the Mandelstam variables. For the scattering amplitude of N particles, there are $N(N-3)/2$ independent variables \cite{ZK} and the rest can be written in terms of the independent ones using conservation of momentum. In the present case there are 15 Mandelstam variables of which 9 are independent. We choose them to be $s_{12}, s_{13},s_{23},s_{24},s_{25},s_{34},s_{35},s_{45},s_{56}$. The restriction \reef{zero} sets $s_{24}=s_{25}=s_{35}=0$. The other six dependent variables can be written in terms of the six non-zero independent variables as
\beqa
s_{14}&=&s_{56}-s_{12}-s_{13}-s_{23}-s_{34}\nonumber\\
s_{15}&=&-s_{45}-s_{56}\nonumber\\
s_{16}&=&s_{23}+s_{34}+s_{45}\nonumber\\
s_{26}&=&-s_{12}-s_{23}\nonumber\\
s_{36}&=&-s_{12}-s_{23}-s_{34}\nonumber\\
s_{46}&=&s_{12}+s_{13}+s_{23}-s_{45}-s_{56}\labell{mandel}
\eeqa
Note that the amplitude has 9 channels. Six of them are $s_{12},s_{23},s_{34},s_{45},s_{56},s_{61}$ and the other three are $(k_1+k_2+k_3)^2,\, (k_2+k_3+k_4)^2$ and $(k_3+k_4+k_5)^2$. The restriction \reef{zero} does not produce singularity in any of these channels.

The correlator in the second line of \reef{amp2} gives different contractions of the scalar polarizations. There are 15  contractions of which 8 involve the contractions \reef{zero1}. The other contractions do not produce terms like $x_{24}^{-n_{24}}x_{25}^{-n_{25}}x_{35}^{-n_{35}}$. 

Let us consider the contraction $\z_1\inn\z_2\,\z_3\inn\z_4\,\z_5\inn\z_6$. The appropriate terms in the second line of \reef{amp2} are
\beqa
-s_{14}(1+s_{56})x_{12}^{-1}x_{34}^{-1}x_{14}^{-1}x_{56}^{-2}+ s_{15}s_{46} x_{12}^{-1}x_{34}^{-1}x_{15}^{-1}x_{56}^{-1}x_{46}^{-1}
- s_{16}s_{45} x_{12}^{-1}x_{34}^{-1}x_{16}^{-1}x_{56}^{-1}x_{45}^{-1}\nonumber
\eeqa
Inserting this into the amplitude \reef{amp2}, one can easily verifies that the integrand is $SL(2,R)$ invariant. Removing this symmetry by fixing $x_1=0,\, x_5=1,\,x_6=\infty$ which has the Jacobian $J\sim x_6^2$, one finds 
\beqa
A_1&\sim&\z_1\inn\z_2\,\z_3\inn\z_4\,\z_5\inn\z_6\Tr(\la_1\la_2\la_3\la_4\la_5\la_6)\int_0^1dx_4\int_0^{x_4}dx_3\int_0^{x_3}dx_2\nonumber\\
&&\bigg(-s_{14}(1+s_{56})x_{12}^{-s_{12}-1}x_{13}^{-s_{13}}x_{14}^{-s_{14}-1}x_{23}^{-s_{23}-1}x_{34}^{-s_{34}-1}x_{45}^{-s_{45}}\nonumber\\
&&+s_{15}s_{46}x_{12}^{-s_{12}-1}x_{13}^{-s_{13}}x_{14}^{-s_{14}}x_{23}^{-s_{23}-1}x_{34}^{-s_{34}-1}x_{45}^{-s_{45}}\nonumber\\
&&-s_{16}s_{45}x_{12}^{-s_{12}-1}x_{13}^{-s_{13}}x_{14}^{-s_{14}}x_{23}^{-s_{23}-1}x_{34}^{-s_{34}-1}x_{45}^{-s_{45}-1}\bigg)\labell{amp21}
\eeqa
Changing the variables $x_2=uvx_4,\, x_3=vx_4$ with the Jacobian $J=vx_4^2$, one can write the above amplitude as
\beqa
A_1&\sim&\z_1\inn\z_2\,\z_3\inn\z_4\,\z_5\inn\z_6\Tr(\la_1\la_2\la_3\la_4\la_5\la_6)\int_0^1dx_4\int_0^1du\int_0^1dv\labell{amp3}\\
&&\bigg(-s_{14}(1+s_{56})x_4^{-s_{56}-2}(1-x_4)^{-s_{45}}u^{-s_{12}-1}(1-u)^{-s_{23}-1}v^{-s_{12}-s_{13}-s_{23}-1}(1-v)^{-s_{34}-1}\nonumber\\
&&+s_{15}s_{46}\,x_4^{-s_{56}-1}(1-x_4)^{-s_{45}}u^{-s_{12}-1}(1-u)^{-s_{23}-1}v^{-s_{12}-s_{13}-s_{23}-1}(1-v)^{-s_{34}-1}\nonumber\\
&&-s_{16}s_{45}\,x_4^{-s_{56}-1}(1-x_4)^{-s_{45}-1}u^{-s_{12}-1}(1-u)^{-s_{23}-1}v^{-s_{12}-s_{13}-s_{23}-1}(1-v)^{-s_{34}-1}\bigg)\nonumber
\eeqa
where we have used the relation $s_{12}+s_{13}+s_{14}+s_{23}+s_{34}=s_{56}$ which can be checked from the relations in \reef{mandel}. Have not restricted the Mandelstam variables to \reef{zero}, one would find term $(x_4(1-uv))^{-s_{24}}(1-uvx_4)^{-s_{25}}(1-vx_4)^{-s_{35}}$ in each line which makes the integral complicated. Using the definition of the beta function
\beqa
\int_0^1dx\, x^{\alpha-1}(1-x)^{\beta-1}&=&B(\alpha,\beta)
\eeqa
one can write the result in terms of three beta functions. The result is
\beqa
A_1&\sim&\z_1\inn\z_2\,\z_3\inn\z_4\,\z_5\inn\z_6\frac{\Ga(-s_{56})\Ga(1-s_{45})\Ga(-s_{12})\Ga(1-s_{23})\Ga(-s_{12}-s_{13}-s_{23})\Ga(-s_{34})}{\Ga(-s_{56}-s_{45})\Ga(-s_{12}-s_{23})\Ga(-s_{12}-s_{13}-s_{23}-s_{34})}\nonumber
\eeqa
where we have also used the relation $s_{14}+s_{16}+s_{46}=s_{23}$. There is also the color factor $\Tr(\la_1\la_2\la_3\la_4\la_5\la_6)$ in the above amplitude.

Doing the same calculation for the other contractions of the polarizations, one finds the following results:
\beqa
A_2&\!\!\!\!\sim\!\!\!\!&\z_1\inn\z_5\,\z_2\inn\z_6\,\z_3\inn\z_4\frac{\Ga(1-s_{56})\Ga(1-s_{45})\Ga(1-s_{12})\Ga(1-s_{23})\Ga(1-s_{12}-s_{13}-s_{23})\Ga(-s_{34})}{\Ga(1-s_{56}-s_{45})\Ga(1-s_{12}-s_{23})\Ga(1-s_{12}-s_{13}-s_{23}-s_{34})}\nonumber\\
A_3&\!\!\!\!\sim\!\!\!\!&\z_1\inn\z_5\,\z_2\inn\z_3\,\z_4\inn\z_6\frac{\Ga(1-s_{56})\Ga(1-s_{45})\Ga(1-s_{12})\Ga(-s_{23})\Ga(-s_{12}-s_{13}-s_{23})\Ga(1-s_{34})}{\Ga(1-s_{56}-s_{45})\Ga(-s_{12}-s_{23})\Ga(1-s_{12}-s_{13}-s_{23}-s_{34})}\nonumber\\
A_4&\!\!\!\!\sim\!\!\!\!&\z_1\inn\z_2\,\z_3\inn\z_6\,\z_4\inn\z_5\frac{\Ga(1-s_{56})\Ga(-s_{45})\Ga(-s_{12})\Ga(1-s_{23})\Ga(-s_{12}-s_{13}-s_{23})\Ga(1-s_{34})}{\Ga(-s_{56}-s_{45})\Ga(-s_{12}-s_{23})\Ga(1-s_{12}-s_{13}-s_{23}-s_{34})}\nonumber\\
A_5&\!\!\!\!\sim\!\!\!\!&\z_1\inn\z_3\,\z_2\inn\z_6\,\z_4\inn\z_5\frac{\Ga(1-s_{56})\Ga(-s_{45})\Ga(1-s_{12})\Ga(1-s_{23})\Ga(-s_{12}-s_{13}-s_{23})\Ga(1-s_{34})}{\Ga(-s_{56}-s_{45})\Ga(1-s_{12}-s_{23})\Ga(1-s_{12}-s_{13}-s_{23}-s_{34})}\nonumber\\
A_6&\!\!\!\!\sim\!\!\!\!&\z_1\inn\z_4\,\z_2\inn\z_3\,\z_5\inn\z_6\frac{\Ga(-s_{56})\Ga(1-s_{45})\Ga(1-s_{12})\Ga(-s_{23})\Ga(-s_{12}-s_{13}-s_{23})\Ga(1-s_{34})}{\Ga(-s_{56}-s_{45})\Ga(-s_{12}-s_{23})\Ga(1-s_{12}-s_{13}-s_{23}-s_{34})}\nonumber\\
A_7&\!\!\!\!\sim\!\!\!\!&\z_1\inn\z_6\,\z_2\inn\z_3\,\z_4\inn\z_5\frac{\Ga(1-s_{56})\Ga(-s_{45})\Ga(1-s_{12})\Ga(-s_{23})\Ga(-s_{12}-s_{13}-s_{23})\Ga(1-s_{34})}{\Ga(-s_{56}-s_{45})\Ga(-s_{12}-s_{23})\Ga(1-s_{12}-s_{13}-s_{23}-s_{34})}\nonumber
\eeqa
Each amplitude includes the color factor $\Tr(\la_1\la_2\la_3\la_4\la_5\la_6)$ as well.

To double check the above results, we examine the symmetry under the cyclic permutation of $(1,2,3,4,5,6)$. From the contraction of the polarizations, one realizes   that the amplitude $A_1$ should be reduced to the amplitude $A_7$ under the permutation $(1,2,3,4,5,6)\rightarrow (6,1,2,3,4,5)$. Since we already restrict the Mandelstam variables to \reef{zero}, we have to set to zero the other Mandelstam variables which map to $s_{24},s_{25},s_{35}$. Under the above permutation one finds
\beqa
(s_{24},s_{25},s_{35}, s_{46},s_{36})\rightarrow (s_{13},s_{14},s_{24},s_{35},s_{25})
\eeqa
So we have to set to zero $s_{13},s_{14},s_{36},s_{46}$ as well. Now again we have to set to zero the  Mandelstam variables which map to $s_{13},s_{14},s_{36},s_{46}$, and so on. In this way we find that to check this symmetry we have to set $s_{13}=s_{14}=s_{46}=s_{36}=s_{26}=s_{15}=0$. This makes  only $s_{56}$ to be nonzero. Using the relations in \reef{mandel}, one can then easily verify that $A_1$ maps to $A_7$ under this permutation.
Similarly $A_2$ maps to $A_3$ under $(1,2,3,4,5,6)\rightarrow (6,1,2,3,4,5)$, $A_2$ maps to $A_5$ under $(1,2,3,4,5,6)\rightarrow (2,3,4,5,6,1))$ and $A_4$ maps to $A_6$ under $(1,2,3,4,5,6)\rightarrow (5,6,1,2,3,4)$.

The  calculation for the other contractions  involves a triplet integral like the one in \reef{amp3} which includes also the factor $(x_4(1-uv))^{-n_{24}}(1-uvx_4)^{-n_{25}}(1-vx_4)^{-n_{35}}$ where at least one of $n_{24},\, n_{25}$ or $n_{35}$ is non-zero. The contractions $\z_1\inn\z_3\z_2\inn\z_4\z_5\inn\z_6$ and $\z_1\inn\z_5\z_2\inn\z_4\z_3\inn\z_6$ include the factor $(x_4(1-uv))^{-1}$. In this case the integral over $dx_4$ gives the  beta function and integral over $dudv$ is of the following form:
\beqa
\int_0^1du\int_0^1dv\,u^av^b(1-u)^c(1-v)^d(1-uv)^{-1}
\eeqa
This integral can be written in terms of Gamma function and Hypergeometric function ${}_3F_{2}$, (see \eg \cite{Oprisa:2005wu})
\beqa
\frac{\Ga(1+a)\Ga(1+b)\Ga(1+c)\Ga(1+d)}{\Ga(2+a+c)\Ga(2+b+d)}{}_3F_2\bigg[{1+a,\ 1+b,\ 1\atop 2+a+c,\ 2+b+d}\ ;\ 1 
\bigg]
 \eeqa 
The result for the above contractions are
\beqa
A_8&\sim&\z_1\inn\z_3\,\z_2\inn\z_4\,\z_5\inn\z_6\Tr(\la_1\la_2\la_3\la_4\la_5\la_6){}_3F_2\bigg[{1-s_{12},\ -s_{12}-s_{13}-s_{23},\ 1\atop 1-s_{12}-s_{23},\ 1-s_{12}-s_{13}-s_{23}-s_{34}}\ ;\ 1 
\bigg]\nonumber\\
&&\times\frac{\Ga(-s_{56})\Ga(1-s_{45})\Ga(1-s_{12})\Ga(1-s_{23})\Ga(-s_{12}-s_{13}-s_{23})\Ga(1-s_{34})}{\Ga(-s_{56}-s_{45})\Ga(1-s_{12}-s_{23})\Ga(1-s_{12}-s_{13}-s_{23}-s_{34})}\nonumber\\
A_9&\sim&\z_1\inn\z_5\,\z_2\inn\z_4\,\z_3\inn\z_6\Tr(\la_1\la_2\la_3\la_4\la_5\la_6){}_3F_2\bigg[{1-s_{12},\ 1-s_{12}-s_{13}-s_{23},\ 1\atop 1-s_{12}-s_{23},\ 2-s_{12}-s_{13}-s_{23}-s_{34}}\ ;\ 1 
\bigg]\nonumber\\
&&\times\frac{\Ga(1-s_{56})\Ga(1-s_{45})\Ga(1-s_{12})\Ga(1-s_{23})\Ga(1-s_{12}-s_{13}-s_{23})\Ga(1-s_{34})}{\Ga(1-s_{56}-s_{45})\Ga(1-s_{12}-s_{23})\Ga(2-s_{12}-s_{13}-s_{23}-s_{34})}\nonumber
\eeqa
where we have only used the relation $x\Ga(x)=\Ga(x+1)$ and $s_{14}+s_{16}+s_{46}=s_{23}$. Note that the above  two terms are not related to each other under the cyclic permutations. Both amplitudes have massless pole in $(k_2+k_3+k_4)^2$-channel which are coming from the pole of the Hypergeometri function.

The contractions $\z_1\inn\z_2\z_3\inn\z_5\z_4\inn\z_6$ and $\z_1\inn\z_4\z_3\inn\z_5\z_2\inn\z_6$ include the factor $(1-x_4v)^{-1}$. In this case the integral over $du$ gives the  beta function and integral over $dx_4dv$ is of the following form:
\beqa
\int_0^1dx_4\int_0^1dv\,x_4^av^b(1-x_4)^c(1-v)^d(1-x_4v)^{-1}
\eeqa
which again can be written in terms of the Gamma function and Hypergeometric function ${}_3F_{2}$. The result in this case is
\beqa
A_{10}&\sim&\z_1\inn\z_2\,\z_3\inn\z_5\,\z_4\inn\z_6\Tr(\la_1\la_2\la_3\la_4\la_5\la_6){}_3F_2\bigg[{1-s_{56},\ -s_{12}-s_{13}-s_{23},\ 1\atop 1-s_{45}-s_{56},\ 1-s_{12}-s_{13}-s_{23}-s_{34}}\ ;\ 1 
\bigg]\nonumber\\
&&\times\frac{\Ga(1-s_{56})\Ga(1-s_{45})\Ga(-s_{12})\Ga(1-s_{23})\Ga(-s_{12}-s_{13}-s_{23})\Ga(1-s_{34})}{\Ga(1-s_{56}-s_{45})\Ga(-s_{12}-s_{23})\Ga(1-s_{12}-s_{13}-s_{23}-s_{34})}\nonumber\\
A_{11}&\sim&\z_1\inn\z_4\,\z_3\inn\z_5\,\z_2\inn\z_6\Tr(\la_1\la_2\la_3\la_4\la_5\la_6) {}_3F_2\bigg[{1-s_{56},\ 1-s_{12}-s_{13}-s_{23},\ 1\atop 1-s_{45}-s_{56},\ 2-s_{12}-s_{13}-s_{23}-s_{34}}\ ;\ 1 
\bigg]\nonumber\\
&&\times\frac{\Ga(1-s_{56})\Ga(1-s_{45})\Ga(1-s_{12})\Ga(1-s_{23})\Ga(1-s_{12}-s_{13}-s_{23})\Ga(1-s_{34})}{\Ga(1-s_{56}-s_{45})\Ga(1-s_{12}-s_{23})\Ga(2-s_{12}-s_{13}-s_{23}-s_{34})}\nonumber
\eeqa
In this case to find the above result we have used the following relation:
\beqa
&&(d-a-1){}_3F_2\bigg[{a,\ b,\ 1\atop c,\ d}\ ;\ 1 \bigg]+(a+b-d-c+2){}_3F_2\bigg[{a+1,\ b,\ 1\atop c,\ d}\ ;\ 1 \bigg]\nonumber\\
&&+\frac{(c-1)(c-b)}{c}{}_3F_2\bigg[{a+1,\ b,\ 1\atop c+1,\ d}\ ;\ 1 \bigg]=0\labell{iden1}
\eeqa
where it can easily be checked for the special case of $b=d$, using the identity ${}_2F_1\bigg[{a,\ 1\atop c}\ ;\ 1 \bigg]=(1-c)/(1+a-c)$. It can also be verified by the Mathematica by expanding the left hand side at $a,b,c,d\rightarrow 0$. One can also find the above result for $A_{10},\, A_{11}$, without using the above identity, if one uses the vertex operators $V(x_4),\, V(x_5)$ in $-1$ picture and the others in $0$ picture. The amplitudes $A_{10}$ and $A_{11}$ are not related to each other under the cyclic permutations. The pole of the Hypergeometric function indicates that both amplitudes have massless pole in $(k_3+k_4+k_5)^2$-channel. The amplitude $A_8$ maps to $A_{10}$ under the cyclic permutation of $(1,2,3,4,5,6)\rightarrow (3,4,5,6,1,2)$, and $A_9$ maps to $A_{11}$ under the cyclic permutation of $(1,2,3,4,5,6)\rightarrow (3,4,5,6,1,2)$.

The contractions $\z_1\inn\z_3\z_2\inn\z_5\z_4\inn\z_6$, $\z_1\inn\z_4\z_2\inn\z_5\z_3\inn\z_6$ and $\z_1\inn\z_6\z_2\inn\z_5\z_3\inn\z_4$ include the factor $(1-uvx_4)^{-1}$.  In this case, the integral over $dx_4dudv$ can be written in terms of the Gamma function and the Hypergeometric function ${}_4F_{3}$, (see \eg \cite{Oprisa:2005wu}). The last contraction $\z_1\inn\z_6\z_2\inn\z_4\z_3\inn\z_5$ includes the factor $(1-vx_4)^{-1}(1-uv)^{-1}$. In this case, the integral over $dx_4dudv$ is even more complicated and  can be written in terms of the triple Hypergeometric function \cite{Oprisa:2005wu}. One can again work out to find the final results which are more complicated than those in $A_1,\cdots, A_{11}$.   We note  that  if one chooses to set to zero three other Mandelstam variables other than those in \reef{zero}, the result would be much easier. For example if one sets $s_{13}=s_{14}=s_{24}=0$ and releases the constraint \reef{zero}, the result for $\z_1\inn\z_3\z_2\inn\z_5\z_4\inn\z_6$ and $\z_1\inn\z_6\z_2\inn\z_5\z_3\inn\z_4$ would be in terms of only Gamma function as in $A_1,\cdots A_{7}$, and the result for  $\z_1\inn\z_6\z_2\inn\z_4\z_3\inn\z_5$ would be in terms of ${}_3F_{2}$ as in $A_{8},\cdots A_{11}$. This would  happened  if one  fixes the $SL(2,R)$ symmetry as $x_{6}=0,\, x_4=1,\, x_5=\infty$.


It would be interesting to study the low energy expansion of the above amplitudes and find the corresponding field theory couplings. In comparing with the  field theory one should, of course, use the same constraint \reef{zero} on the Mandelstam variables. This makes the study much more easier than the more general case \cite{Medina:2002nk, Medina:2006uf,Oprisa:2005wu}. This study helps one to find $(D\phi)^6$ terms of the effective action, hence, using the T-duality transformation $(D\phi)^6\rightarrow F^6$, one would be able to find  the corrections to the symmetrized trace nonabelian DBI action \cite{Tseytlin:1997csa,Myers:1999ps}. We postpone this study to the future works and focus here in the next section to the ER limit of these amplitudes. 

\subsection{ The ER limit}

The adjacent BCFW shift that is consistent with the constraint \reef{zero} is
 \beqa
k_6&\rightarrow &\hat{k}_6=k_6+qz\nonumber\\
k_{1}&\rightarrow &\hat{k}_{1}=k_{1}-qz
\eeqa
where $q$ satisfies $q\inn q=k_{1}\inn q=k_{6}\inn q=0$. In the ER limit,  one finds $s_{12},s_{13},s_{56}$ are large and $s_{23},s_{34},s_{45},s_{24},s_{25},s_{35}$ are small. 

To study  the amplitudes $A_1,\cdots A_7$ in this limit, we write them as
\beqa
A_1&\!\!\!\!\sim\!\!\!\!&\z_1\inn\z_2\,\z_3\inn\z_4\,\z_5\inn\z_6\Tr(\la_1\la_2\la_3\la_4\la_5\la_6)\frac{(-s_{45}-s_{56})(-s_{12}-s_{23})(-s_{12}-s_{13}-s_{23}-s_{34})}{(-s_{56})(-s_{12})(-s_{12}-s_{13}-s_{23})(-s_{34})}F\nonumber\\
A_2&\!\!\!\!\sim\!\!\!\!&\z_1\inn\z_5\,\z_2\inn\z_6\,\z_3\inn\z_4\Tr(\la_1\la_2\la_3\la_4\la_5\la_6)\frac{1}{(-s_{34})}F\nonumber\\
A_3&\!\!\!\!\sim\!\!\!\!&\z_1\inn\z_5\,\z_2\inn\z_3\,\z_4\inn\z_6\Tr(\la_1\la_2\la_3\la_4\la_5\la_6)\frac{(-s_{12}-s_{23})}{(-s_{23})(-s_{12}-s_{13}-s_{23})}F\nonumber\\
A_4&\!\!\!\!\sim\!\!\!\!&\z_1\inn\z_2\,\z_3\inn\z_6\,\z_4\inn\z_5\Tr(\la_1\la_2\la_3\la_4\la_5\la_6)\frac{(-s_{45}-s_{56})(-s_{12}-s_{23})}{(-s_{45})(-s_{12})(-s_{12}-s_{13}-s_{23})}F\nonumber\\
A_5&\!\!\!\!\sim\!\!\!\!&\z_1\inn\z_3\,\z_2\inn\z_6\,\z_4\inn\z_5\Tr(\la_1\la_2\la_3\la_4\la_5\la_6)\frac{(-s_{45}-s_{56})}{(-s_{45})(-s_{12}-s_{13}-s_{23})}F\nonumber\\
A_6&\!\!\!\!\sim\!\!\!\!&\z_1\inn\z_4\,\z_2\inn\z_3\,\z_5\inn\z_6\Tr(\la_1\la_2\la_3\la_4\la_5\la_6)\frac{(-s_{45}-s_{56})(-s_{12}-s_{13})}{(-s_{56})(-s_{23})(-s_{12}-s_{13}-s_{23})}F\nonumber\\
A_7&\!\!\!\!\sim\!\!\!\!&\z_1\inn\z_6\,\z_2\inn\z_3\,\z_4\inn\z_5\Tr(\la_1\la_2\la_3\la_4\la_5\la_6)\frac{(-s_{45}-s_{56})(-s_{12}-s_{23})}{(-s_{45})(-s_{23})(-s_{12}-s_{13}-s_{23})}F\labell{ampf}
\eeqa
where the string form factor $F$ includes the Gamma functions
\beqa
F&=&\frac{\Ga(1-s_{56})\Ga(1-s_{45})\Ga(1-s_{12})\Ga(1-s_{23})\Ga(1-s_{12}-s_{13}-s_{23})\Ga(1-s_{34})}{\Ga(1-s_{56}-s_{45})\Ga(1-s_{12}-s_{23})\Ga(1-s_{12}-s_{13}-s_{23}-s_{34})}\labell{form}
\eeqa
At low energy region, $s_{ij}\rightarrow  0$,  the factor $F$  reduces  to one and the rest should be reproduced by the field theory coupling $\Tr(D_{\mu}\phi^iD^{\mu}\phi^i)$. We have checked explicitly that for example $ A_2$ is reproduced by the following Feynman amplitude:
\beqa
V(612\phi)G(\phi)V(\phi 345)+V(612\phi)G(\phi)V(\phi 5 A)G(A)V(A34)\nonumber
\eeqa
where our notation is such that  $V(612\phi)$ is the vertex for four scalars in which the particles 6,1,2 are on-shell and the particle $\phi$ is off-shell, and $G(\phi)$ is the propagators of the scalar field.

At the ER limit, $s_{12},s_{13},s_{56}\rightarrow\infty$ and $s_{23},s_{34},s_{45}\rightarrow 0$, so the string form factor $F$ again reduces to one.

To analyze the large $z$ behavior of the amplitudes $A_8,\,A_9$, we note that these amplitudes have no massless pole in $s_{23}$-channel. So to simplify the discussion we set $s_{23}=0$. In this case these amplitudes can be written as
\beqa
A_8&\sim&\z_1\inn\z_3\,\z_2\inn\z_4\,\z_5\inn\z_6\Tr(\la_1\la_2\la_3\la_4\la_5\la_6)\frac{(-s_{56}-s_{45})(-s_{12}-s_{13}-s_{34})}{(-s_{56})(-s_{34})(-s_{12}-s_{13})}F\nonumber\\
A_9&\sim&\z_1\inn\z_5\,\z_2\inn\z_4\,\z_3\inn\z_6\Tr(\la_1\la_2\la_3\la_4\la_5\la_6)\frac{1}{(-s_{34})}F\nonumber
\eeqa
where $F$ is the same form factor \reef{form} in which $s_{23}=0$. Note that the massless pole in $s_{34}$-channel in the above amplitudes is in fact the massless pole in $(k_2+k_3+k_4)^2$-channel. The form factor reduces to one at the  ER limit and the rest are the  amplitudes which are reproduced by the nonabelian kinetic term of the scalar field.

The amplitudes $A_{10},A_{11}$ have no massless pole in $s_{45}$-channel. So it is consistent to set $s_{45}=0$. In this case they simplify to
\beqa
A_{10}&\sim&\z_1\inn\z_2\,\z_3\inn\z_5\,\z_4\inn\z_6\Tr(\la_1\la_2\la_3\la_4\la_5\la_6)\frac{(-s_{12}-s_{23})(-s_{12}-s_{13}-s_{23}-s_{34})}{(-s_{12})(-s_{34})(-s_{12}-s_{13}-s_{23})}F\nonumber\\
A_{11}&\sim&\z_1\inn\z_4\,\z_3\inn\z_5\,\z_2\inn\z_6\Tr(\la_1\la_2\la_3\la_4\la_5\la_6)\frac{1}{(-s_{34})}F\nonumber
\eeqa
where $F$ is the same form factor \reef{form} in which $s_{45}=0$. Here again  the massless pole in $s_{34}$-channel  is in fact the massless pole in $(k_3+k_4+k_5)^2$-channel. The form factor reduces to one at the  ER limit and the rest are the  amplitudes which are reproduced by the field theory.

The string form factors in the amplitudes $A_{12},\cdots, A_{15}$ are more complicated for the constraint \reef{zero}. However, if one uses another constraint the result would be much simpler, as in $A_{1},\cdots, A_{11}$. We note that in other constraint the adjacent BCFW shift are changed and the form factors are again reduce to one at the ER limit, \eg  the adjacent BCFW shift  for the constraint $s_{13}=s_{14}=s_{24}=0$ is $\hat{k}_5,\, \hat{k}_6$.

\section{Eight-point functions}

We have done the calculation for the six-point function in the case that the vertex operators $V(x_2), V(x_3)$ are in -1 picture and the rest are in 0 picture. This makes the calculation to have no terms proportional to $s_{24},\, s_{25}$ or $s_{35}$ as in \reef{amp21}. To simplify this calculation we note that the amplitude 
has no channel in $s_{24},\, s_{25}$ or $s_{35}$.  So one can use another arrangement for the vertex operators in which some of the correlators are proportional to $s_{24},\, s_{25}$ or $s_{35}$. They are then zero under the constraint \reef{zero}.  The arrangement in which $V(x_1),\, V(x_6)$ are in -1 picture  only one correlator  survives the constraint \reef{zero},  instead of three terms in \reef{amp21}. The calculation becomes much more easier to perform, and the final result is unchanged. Hence, for the eight-point functions we choose the vertex operators $V(x_1),\, V(x_8)$ to be in -1 picture and the rest in 0 picture, that is  
\beqa
A&\sim &\int dx_1\cdots dx_8<V^{-1}(x_1)V^0(x_2)\cdots V^0(x_7)V^{-1}(x_8)>\Tr(\la_1\la_2\la_3\la_4\la_5\la_6\la_7\la_8)\nonumber
\eeqa
The position of the vertices are $-\infty<x_1<x_2<x_3<x_4<x_5<x_6<x_7<x_8<\infty$. 

There are 20 independent Mandelstam variables in the eight-point function. We restrict the amplitude to the case that 
\beqa
s_{24}=s_{25}=s_{26}=s_{27}=s_{35}=s_{36}=s_{37}=s_{46}=s_{47}=s_{57}=0\labell{zero8}
\eeqa
Note that the above constraint does not change the number of channels.  There are 8 adjacent channels in  $s_{12},s_{23},s_{34},s_{45},s_{56},s_{67},s_{78},s_{81}$,  8 channels  in $(k_1+k_2+k_3)^2$, $(k_2+k_3+k_4)^2$, $(k_3+k_4+k_5)^2$,$(k_4+k_5+k_6)^2$, $(k_5+k_6+k_7)^2$, $(k_6+k_7+k_8)^2$, $(k_7+k_8+k_1)^2$, $(k_8+k_1+k_2)^2$, and the other 4 channels are in $(k_1+k_2+k_3+k_4)^2$, $(k_2+k_3+k_4+k_5)^2$, $(k_3+k_4+k_5+k_6)^2$, $(k_4+k_5+k_6+k_7)^2$. The above restriction does no produce any singularity in these channels.

There are 105 different contractions of the scalar polarization. Some of them can be written in terms of only Gamma functions as in $A_1,\cdots, A_7$ and the rest in terms of the Gamma and some more complicated functions. However, there are always a particular constraint as in \reef{zero8} in which  a given contraction can be written in terms of only Gamma functions. In the constraint \reef{zero8}, the contractions of the scalar polarizations  which do not include 
\beqa
\z_2\inn\z_4, \z_2\inn\z_5,\z_2\inn\z_6,\z_2\inn\z_7, \z_3\inn\z_5,\z_3\inn\z_6,\z_3\inn\z_7, \z_4\inn\z_6,\z_4\inn\z_7, \z_5\inn\z_7
\eeqa
can be written in terms of only the Gamma functions. Since there are many terms of this type,  we only consider the adjacent  contraction, \ie $\z_1\inn\z_2\z_3\inn\z_4\z_5\inn\z_6\z_7\inn\z_8$.

Using the constraint \reef{zero8} there is only one nonzero term. The integrand  is invariant under the $SL(2,R)$. Removing this symmetry by fixing $x_1=0,\, x_7=1$ and $x_8=\infty$, one finds 
\beqa
{\cal A}&\sim&\z_1\inn\z_2\z_3\inn\z_4\z_5\inn\z_6\z_7\inn\z_8\Tr(\la_1\la_2\la_3\la_4\la_5\la_6\la_7\la_8)\int_0^1dx_6\int_0^{x_6}dx_5\int_0^{x_5}dx_4\int_0^{x_4}dx_3\int_0^{x_3}dx_2\nonumber\\
&&s_{23}s_{45}s_{67}(x_{12}^{-s_{12}-1}x_{13}^{-s_{13}}x_{14}^{-s_{14}}x_{15}^{-s_{15}-1}x_{16}^{-s_{16}-1}x_{23}^{-s_{23}-1}x_{34}^{-s_{34}-1}x_{45}^{-s_{45}-1}x_{56}^{-s_{56}-1}x_{67}^{s_{67}-1})\nonumber
\eeqa
We take the Mandelstam variables that appear in the above equation, \ie $s_{12}, s_{13}, s_{14}, s_{15}, s_{16}$, $s_{23}, s_{34}, s_{45},s_{56}, s_{67}$ as the independent variables. Note that these 10 variables and the 10 variables in the constraint \reef{zero8} make the total Mandelstam variables of the eight-point function. Changing the variables as $x_5=\alpha x_6,\, x_4=\alpha\beta x_6,\, x_3=\alpha\beta u x_6$, and $x_2=\alpha\beta uv x_6$ which has the Jacobian $J=u\beta^2\alpha^3 x_6^4$, one finds that the four integrals are separated and each one can be written in terms of the beta function. The result is
\beqa
{\cal A}&\sim&\z_1\inn\z_2\z_3\inn\z_4\z_5\inn\z_6\z_7\inn\z_8\Tr(\la_1\la_2\la_3\la_4\la_5\la_6\la_7\la_8)\labell{ampf8}\\
&&\times\frac{\Ga(-s_{12}-s_{13}-s_{14}-s_{15}-s_{16}-s_{23}-s_{34}-s_{45}-s_{56})\Ga(1-s_{67})}{\Ga(-s_{12}-s_{13}-s_{14}-s_{15}-s_{16}-s_{23}-s_{34}-s_{45}-s_{56}-s_{67})}\nonumber\\
&&\times\frac{\Ga(-s_{12}-s_{13}-s_{14}-s_{15}-s_{23}-s_{34}-s_{45})\Ga(-s_{56})}{\Ga(-s_{12}-s_{13}-s_{14}-s_{15}-s_{23}-s_{34}-s_{45}-s_{56})}\nonumber\\
&&\times\frac{\Ga(-s_{12}-s_{13}-s_{14}-s_{23}-s_{34})\Ga(1-s_{45})}{\Ga(-s_{12}-s_{13}-s_{14}-s_{23}-s_{34}-s_{45})}\nonumber\\
&&\times\frac{\Ga(-s_{12}-s_{13}-s_{23})\Ga(-s_{34})}{\Ga(-s_{12}-s_{13}-s_{23}-s_{34})}\frac{\Ga(-s_{12})\Ga(1-s_{23})}{\Ga(-s_{12}-s_{23})}\nonumber
\eeqa
Note that using the first relation in \reef{mandel}, one observes that $A_1$ in the previous section has a structure as above. One may try to find a similar result for the other contractions of the polarizations, as in section 3. 

To study the above amplitude at the ER limit we write the amplitude as the following:
\beqa
{\cal A}&\sim& {\cal A}_{YM}F
\eeqa
where the field theory amplitude is
\beqa
{\cal A}_{YM}&=&\z_1\inn\z_2\z_3\inn\z_4\z_5\inn\z_6\z_7\inn\z_8\Tr(\la_1\la_2\la_3\la_4\la_5\la_6\la_7\la_8)\labell{ampf81}\\
&&\times\frac{(-s_{12}-s_{13}-s_{14}-s_{15}-s_{16}-s_{23}-s_{34}-s_{45}-s_{56}-s_{67})}{(-s_{12}-s_{13}-s_{14}-s_{15}-s_{16}-s_{23}-s_{34}-s_{45}-s_{56})}\nonumber\\
&&\times\frac{(-s_{12}-s_{13}-s_{14}-s_{15}-s_{23}-s_{34}-s_{45}-s_{56})}{(-s_{12}-s_{13}-s_{14}-s_{15}-s_{23}-s_{34}-s_{45})(-s_{56})}\nonumber\\
&&\times\frac{(-s_{12}-s_{13}-s_{14}-s_{23}-s_{34}-s_{45})}{(-s_{12}-s_{13}-s_{14}-s_{23}-s_{34})}\nonumber\\
&&\times\frac{(-s_{12}-s_{13}-s_{23}-s_{34})}{(-s_{12}-s_{13}-s_{23})(-s_{34})}\frac{(-s_{12}-s_{23})}{(-s_{12})}\nonumber
\eeqa
and the string form factor is 
\beqa
F&=&\frac{\Ga(1-s_{12}-s_{13}-s_{14}-s_{15}-s_{16}-s_{23}-s_{34}-s_{45}-s_{56})\Ga(1-s_{67})}{\Ga(1-s_{12}-s_{13}-s_{14}-s_{15}-s_{16}-s_{23}-s_{34}-s_{45}-s_{56}-s_{67})}\nonumber\\
&&\times\frac{\Ga(1-s_{12}-s_{13}-s_{14}-s_{15}-s_{23}-s_{34}-s_{45})\Ga(1-s_{56})}{\Ga(1-s_{12}-s_{13}-s_{14}-s_{15}-s_{23}-s_{34}-s_{45}-s_{56})}\nonumber\\
&&\times\frac{\Ga(1-s_{12}-s_{13}-s_{14}-s_{23}-s_{34})\Ga(1-s_{45})}{\Ga(1-s_{12}-s_{13}-s_{14}-s_{23}-s_{34}-s_{45})}\nonumber\\
&&\times\frac{\Ga(1-s_{12}-s_{13}-s_{23})\Ga(1-s_{34})}{\Ga(1-s_{12}-s_{13}-s_{23}-s_{34})}\frac{\Ga(1-s_{12})\Ga(1-s_{23})}{\Ga(1-s_{12}-s_{23})}\nonumber
\eeqa
 The adjacent BCFW shift which is consistent with the constraint \reef{zero8} is  
 \beqa
k_8&\rightarrow &\hat{k}_8=k_8+qz\nonumber\\
k_{1}&\rightarrow &\hat{k}_{1}=k_{1}-qz
\eeqa
At the ER limit $s_{23}, s_{34}, s_{45},s_{56}, s_{67}\rightarrow 0$. Hence, the form factor reduces to one at the ER limit.

\section{2n-point functions}

One can easily extend the above calculation  to the general case of 2n-point functions. The amplitude is given by the following correlation function:
\beqa
A&\sim &\int dx_1\cdots dx_{2n}<V^{-1}(x_1)V^0(x_2)\cdots V^0(x_{2n-1})V^{-1}(x_{2n})>\Tr(\la_1\la_2\cdots \la_{2n})\nonumber
\eeqa
The appropriate constraint is
\beqa
\pmatrix{s_{24}& s_{25}& s_{26}&\cdots &s_{2,2n-1}\cr 
&s_{35}& s_{36}& \cdots& s_{3,2n-1}\cr
&& s_{46}& \cdots&s_{4,2n-1}\cr
&&& \vdots& \vdots\cr
&&&& s_{2n-3,2n-1}}=0
 \labell{zero2n} \eeqa 
or any other set under cyclic permutation of $(1,2,\cdots ,2n)$. This makes $(2n-3)(2n-4)/2$ out of  the total $2n(2n-3)/2$ independent Mandelstam variables to be zero. The remaining $2(2n-3)$ variables are chosen to be 
\beqa
s_{12},s_{13},\cdots s_{1,2n-2}, s_{23},s_{34}, \cdots s_{2n-2,2n-1}
\eeqa
There is no channel in the $s_{ij}$'s in the constraint \reef{zero2n}, so the multiple integral has no singularity after imposing the constraint \reef{zero2n}. The $SL(2,R)$ symmetry of the integrand is fixed as $x_1=0,\, x_{2n-1}=1$ and $x_{2n}=\infty$. The amplitude for the contractions of the scalar polarizations which do not include the following contractions:
\beqa
\pmatrix{\z_2\inn \z_4& \z_2\inn \z_5& \z_2\inn \z_6&\cdots &\z_2\inn \z_{2n-1}\cr 
&\z_3\inn \z_5& \z_3\inn \z_6& \cdots& \z_3\inn \z_{2n-1}\cr
&& \z_4\inn \z_6& \cdots&\z_4\inn \z_{2n-1}\cr
&&& \vdots& \vdots\cr
&&&& \z_{2n-3}\inn \z_{2n-1}}
 \labell{zero2nz} \eeqa 
can be written in terms of only the Gamma functions. The result for  the adjacent contractions of the scalar polarizations is
\beqa
{\cal A}&\sim& {\cal A}_{YM}F
\eeqa
where the field theory amplitude is
\beqa
{\cal A}_{YM}&=&\Tr(\la_1\la_2\cdots \la_{2n})\prod_{j=2}^{2n}\z_{2j-3}\inn \z_{2j-2}\frac{(-s_{12}-\cdots -s_{1,2j-2}-s_{23}-\cdots -s_{2j-2,2j-1})}{(-s_{12}-\cdots -s_{1,2j-2}-s_{23}-\cdots -s_{2j-3,2j-2})}\nonumber\\
&&\times\frac{(-s_{12}-\cdots -s_{1,2j-3}-s_{23}-\cdots -s_{2j-3,2j-2})}{(-s_{12}-\cdots -s_{1,2j-3}-s_{23}-\cdots -s_{2j-4,2j-3})(-s_{2j-3,2j-2})}
\eeqa
and the string form factor is
\beqa
F&=&\prod_{j=2}^{2n}\frac{\Ga(1-s_{12}-\cdots -s_{1,2j-2}-s_{23}-\cdots -s_{2j-3,2j-2})\Ga(1-s_{2j-2,2j-1})}{\Ga(1-s_{12}-\cdots -s_{1,2j-2}-s_{23}-\cdots -s_{2j-2,2j-1})}\nonumber\\
&&\times\frac{\Ga(1-s_{12}-\cdots -s_{1,2j-3}-s_{23}-\cdots -s_{2j-4,2j-3})\Ga(1-s_{2j-3,2j-2})}{\Ga(1-s_{12}-\cdots -s_{1,2j-3}-s_{23}-\cdots -s_{2j-3,2j-2})}
\eeqa
The adjacent BCFW shift which is consistent with the constraint \reef{zero2n} is  
 \beqa
k_{2n}&\rightarrow &\hat{k}_{2n}=k_{2n}+qz\nonumber\\
k_{1}&\rightarrow &\hat{k}_{1}=k_{1}-qz
\eeqa
At the ER limit $s_{23},s_{34}, \cdots s_{2n-2,2n-1}\rightarrow 0$. Hence, the form factor reduces to one at the ER limit.

There are many other contraction of the scalar polarizations which can be written in a closed form in terms of the Gamma functions as in section 3 for six-point functions. By explicit calculation, it should be easy to show that the form factor for all of them reduce to one in the ER limit, as in section 3. For those contractions that their form factor involves more complicated functions, one may choose another set of constraint, instead of \reef{zero2n}. The form factors would then be in terms of  only   the Gamma functions. Hence, the form factor for all contractions of the scalar polarizations should reduce to one  at the ER limit. We expect similar discussion should be valid  in the scattering amplitude of gluons in which the gluon polarizations contract    with momentum and with the other gluon polarizations.

{\bf Acknowledgments}:  This work is supported by Ferdowsi University of Mashhad. 
\bibliographystyle{/Users/Nick/utphys} 
\bibliographystyle{utphys} \bibliography{hyperrefs-final}


\providecommand{\href}[2]{#2}\begingroup\raggedright

\endgroup

\end{document}